\documentclass[11pt]{article}
\usepackage{epsfig}
\usepackage{amsmath}
\usepackage{amssymb}
\usepackage{geometry}
\usepackage{url}

\newcommand{\xf}[1]{Figure~\ref{#1}}
\newcommand{\xt}[1]{Table~\ref{#1}}

\newcommand{\xs}[1]{Section~\ref{#1}}

\newtheorem{theorem}{Theorem}
\newtheorem{lemma}{Lemma}

\newtheorem{defn}{Definition}

\def\db{\mbox {$\cal D$}}

\def\tdb{\mbox{$t(\db_{\alpha})$}}
\def\tT{\mbox{$t(T_{\alpha})$}}
\def\tT{\mbox{$t(T_{\alpha})$}}
\def\nT{\mbox{$\nu(T_{\alpha})$}}
\def\njT{\mbox{$\nu[j](T_{\alpha})$}}
\def\mjT{\mbox{$\mu[j](T_{\alpha})$}}

\newcommand{\negvs}{\vspace*{-1ex}}		

\newlength{\qedlengte}
\newcommand{\qedbox}{\rule{\qedlengte}{\qedlengte}}
\newcommand{\qed}{\hspace*{1em}\hfill\qedbox}

\title{Mining Frequent Itemsets from Secondary Memory}

\author{G\"{o}sta Grahne and Jianfei Zhu\\
Concordia University\\
Montreal, Canada\\
\{grahne, j\_zhu\}@cs.concordia.ca\\
}
\date{March 6, 2004}
\begin{document}
\maketitle

\begin{abstract}
Mining frequent itemsets is at the core
of mining association rules, and is by now quite well
understood algorithmically.
However, most algorithms for mining frequent
itemsets assume that the main memory is large enough
for the data structures used in the mining,
and very few efficient algorithms deal with the case when
the database is {\em very} large or the minimum support is very low.
Mining frequent itemsets from a very large database
poses new challenges,
as astronomical amounts of raw data
is ubiquitously being recorded in commerce, science and government.

In this paper, we discuss approaches to mining frequent itemsets when
data structures are too large to fit in main memory.
Several
divide-and-conquer
algorithms
are given for mining from disks.
Many novel techniques are introduced.
Experimental results show that
the techniques reduce
the required disk accesses by orders of magnitude,
and enable truly scalable data mining.
\end{abstract}

\section{Introduction}

Mining frequent itemsets is a fundamental problem 
for mining association rules \cite{AIS93, AS94,MTV94,PBT99, PHM00,WHP03,Zaki02, ZB03}.
It also plays an important role in many other data mining tasks
such as sequential patterns, episodes, multi-dimensional patterns and so on
\cite{AS95, MTV97, KHC97}.
In addition, frequent itemsets are one of the key abstractions
in data mining.

The description of the problem is as follows. 
Let $I = \{i_1,i_2,\ldots,i_j,\ldots i_n\}$,
be a set of {\em items}.
Items will sometimes also be denoted 
$a,b,c,\ldots$.
An $I$-{\em transaction} $\tau$ is a subset of $I$.
An $I$-transactional {\em database} $\db$ is a finite bag
of $I$-transactions.
The {\em support} of an itemset $S\subseteq I$
is the proportion of transactions in \db~ that contain $S$.
The task of mining frequent itemsets is to find 
all $S$ such that the support of $S$ is greater than some 
given {\em minimum support} $\xi$,
where $\xi$ either is a fraction in $[0,1]$,
or an absolute count.

Most of the algorithms, such as 
Apriori \cite{AS94},
DepthProject \cite{AAP00},
and dEclat \cite{Zaki03}
work well when the main memory is big enough to
fit the whole database or/and the data structures
(candidate sets, FP-trees, etc).
When a database is very large or when the minimum support is very low,
either the data structures used by the algorithms may not be accommodated in 
main memory, 
or the algorithms spend too much time on 
multiple passes over the database.
In the 
{\em First IEEE ICDM Workshop on Frequent Itemset 
Mining Implementations, FIMI~'03} \cite{ZB03},
many well known algorithms were implemented
and independently tested.
The results show that ``{\em none} of the algorithms is able to gracefully
scale-up to very large datasets, 
with millions of transactions''
\cite{ZaBa03}.

At the same time
very large databases do exist in real life. 
In a medium sized business or in a company big as Walmart, 
it's very easy to collect a few gigabytes of data. 
Terabytes of raw data
is ubiquitously being recorded in commerce, science and government.
The question of how to handle these databases is still one of the most 
difficult problems in data mining.

A few researchers have
tried to mine frequent itemsets from very large databases.
One approach is by {\em sampling}.
For instance, \cite{Toiv96}
picks a random sample of the database,
finds all frequent itemsets from the sample, and then verifies 
the results with the rest of the database. 
This approach needs only one pass of the database.
However, the results are probabilistic, 
meaning that
some critical frequent itemsets could be missing.

{\em Partitioning} \cite{SON95} 
is another approach for mining very large databases.
This approach first partitions the database 
into many small databases,
and mines candidate frequent itemsets from each small database.
One more pass 
over the original database 
is then done to verify the candidate frequent itemsets.
The approach thus needs only two database scans. 
However, when the data structures used for storing 
candidate frequent itemsets
are too big to fit in main memory, 
a significant amount of  disk I/O's is needed 
for the disk resident data structures. 

In \cite{HPY00, HPYM04}, Han {\em et.\ al.} introduce the  {\em FP-growth} 
method, which 
uses two database scans for constructing an FP-tree
from the database, 
and then mines all frequent itemsets from the FP-tree.
Two approaches are suggested for the case that 
the FP-tree is too large to fit into main memory.

The first approach writes the FP-tree to disk,
then mines all frequent sets by reading the
frequency information from the FP-tree.
However, the size of the FP-tree could be same as the 
size of the database, and for each item in the FP-tree,
we need at least one FP-tree traversal. 
Thus the I/O's for writing and reading the 
disk-resident FP-tree could be
prohibitive.

The second approach
{\em projects} the original database
on each frequent item, then mines frequent itemsets from 
the small projected databases. 
One advantage of this approach is that any frequent itemset 
mined from a projected database is a frequent itemset in the original database.
To get {\em all} frequent itemsets, 
we only need to 
take the union of the frequent itemsets from the small projected databases.
This is in contrast to the
partitioning approach, 
where all candidate frequent itemsets have to be stored and later verified
by another pass of database. 
The biggest problem of the projection approach is that
the total size of the projected databases could be too large, 
and there will be too many disk I/O's for the 
projected databases.

\subsubsection*{Contributions} 
In this paper we consider the problem of  mining frequent itemsets 
from {\em very} large databases. 
We adopt a 
divide-and-conquer approach.
First we give  three algorithms, 
the general divide-and-conquer algorithm, 
then an algorithm using
naive projection, and an algorithm using
aggressive projection.
We also analyze the 
number of steps and disk I/O's required by these algorithms. 

In a detailed divide-and-conquer algorithm,
called {\em Diskmine}, 
we use the highly efficient 
{\em FP-growth*} method \cite{fimi03} to
mine frequent itemsets from an FP-tree in main memory.
We describe several novel techniques
useful in mining frequent itemsets from disks, 
such as the array technique,
the item-grouping technique, 
and memory management techniques.

Finally, we present experimental results that 
demonstrate the fact that our {\em Diskmine}-algorithm
outperforms previous algorithms
by orders of magnitude,
and scales up to terabytes of data.

\subsubsection*{Overview} 
The remainder of this paper is organized as follows. 
In Section 2
we introduce approaches for mining frequent itemsets from disks.
Three algorithms are introduced and analyzed. 
Section 3 gives a detailed divide-and-conquer 
algorithm {\em Diskmine}, 
in which many novel optimization techniques are used. 
These techniques are also described in Section 3.
Experimental results are given in Section 4.
Section 5 concludes, 
and outlines directions for future research.

\section{Mining from disk} \label{diskmine}

How should one go about when mining
frequent itemsets from very large databases
residing in a secondary memory storage,
such as disks?
Here ``very large'' means that 
the data structures constructed from the database 
for mining frequent itemsets
can not fit in the available main memory.

Basically, there are two strategies  
for mining frequent itemsets,
the datastructures approach,
and the 
divide-and-conquer approach.

The {\em datastructures} approach consists of
reading 
the database buffer by buffer, 
and generate
datastructures (i.e.\ candidate sets or FP-trees).
Since the datastructure don't fit into main memory, 
additional disk I/O's are required.
The number of passes and disk I/O's required
by the approach
depends on the algorithm and its datastructures.
For example,
if the algorithm is Apriori \cite{AS94}
using a hash-tree 
for candidate itemsets
\cite{SON95},
disk based hash-trees have to be used.
Then the number of passes for the algorithm
is same as the length of the longest 
frequent itemset,
and the number of disk I/O's for the hash-trees
depend on the size of the hash-trees 
on disk.

The basic strategy for the  
{\em divide-and-conquer} approach
is shown in \xf{bdaqalgo}.
In the approach,
$|\db|$ denotes   
the size of the data structures used 
by the mining algorithm, and
$M$ is the size of available main memory.
Function {\em mainmine}
is called if 
candidate frequent itemsets (not necessary all)
can be mined without 
writing the data structures used by 
a mining algorithm  to disks.
In \xf{bdaqalgo},
a very large database is decomposed into a number 
of smaller databases.
If a ``small'' database is still too large,
i.e, the data structures are still too big to fit in main memory,
the decomposition is recursively continued
until
the data structures fit in main memory. 
After all small databases are processed, 
all candidate frequent itemsets are combined in some way
(obviously depending on the way the decomposition was done)
to get all frequent itemsets for the original database.

\begin{figure}[h]
{\bf Procedure} {\em diskmine}($\db,M$)

\smallskip

{\bf if} $|\db|\leq M$ {\bf then} {\bf return} {\em mainmine($\,\db$)} 

{\bf else} decompose $\db$ into $\db_1,\ldots \db_k$.

{\hskip 18pt}
{\bf return}  {\em combine} {\em diskmine($\,\db_1,M$)},

{\hskip 154pt}                           ....              ,

{\hskip 90pt}       {\em diskmine($\,\db_k,M$)}.

\caption{{\small 
General divide-and-conquer algorithm for 
mining frequent itemsets from disk.
}}
\label{bdaqalgo}
\end{figure}

The efficiency of {\em diskmine} 
depends on 
the method used for mining frequent itemsets
in main memory and on the number of 
disk I/O's needed in the decomposition and
combination phases. 
Sometimes the disk I/O is the main factor.
Since the decomposition step involves I/O,
ideally the number of recursive calls should be
kept small. The faster we can obtain small decomposed
databases, the fewer recursive call we will need.
On the other hand, if a decomposition cuts
down the size of the projected databases drastically, 
the trade-off might be that the combination
step becomes more complicated and might involve heavy 
disk I/O.

In the following we discuss two decomposition
strategies, namely
decomposition by partition, and
decomposition by projection.

{\em Partitioning} 
is an approach in which a large database is decomposed into 
cells of small non-overlapping databases. 
The cell-size is chosen so that
all frequent itemsets in a cell can be mined without 
having to store any data structures in  secondary memory.
However, since a cell only contains partial frequency 
information of the original database,
all frequent itemsets from the cell are local
to that cell of the partition,
and could only be {\em candidate} frequent itemsets
for the whole database.
Thus the candidate frequent itemsets mined from
a cell
have to be verified
later to filter out false hits.
Consequently, 
those candidate sets have to  be written to disk
in order to leave
space for processing the next cell of the partition.
After generating candidate frequent itemsets from 
all cells,
another database scan is needed to 
filter out all infrequent itemsets.
The partition approach therefore needs only two passes
over the database,
but writing and reading candidate frequent itemsets
will involve a significant number of
disk I/O's,
depending on the size of the set of candidate frequent itemsets.

We can conclude that the partition approach 
to decomposition keeps the recursive levels
down to one, but the penalty is that the 
combination phase becomes expensive.

To get an easier combination phase,
we adopt another decomposition strategy, which we call
{\em projection}.
Suppose for simplicity that there are four
items, $a,b,c,$ and $d$, and let $\db$ be a
database of transactions containing some 
or all of these items.
We could then decompose
$\db$ into for instance
$\db_{ab}$ and
$\db_{cd}$.
Typically, we would do this when the descending order
of frequency of the items is $a, b, c, d$.
In $\db_{cd}$ we put all transactions
containing at $c$ or $d$ (or both).
In $\db_{ab}$ we put transactions containing
$a$ or $b$ (or both), and for each transaction we store
only the $a,b$-part. Thus we will have shorter
transactions in $\db_{ab}$, and both
$\db_{ab}$ and
$\db_{cd}$ contain fewer transactions than $\db$.
We can then recursively mine all frequent itemsets
from $\db_{ab}$, and $\db_{cd}$.
Since this decomposition is not a partition,
the projected databases
might not be that much smaller that the
original database. The upside is though that
the set of all frequent itemsets in
$\db$ now simply is the union of the frequent
itemsets in $\db_{ab}$ and $\db_{cd}$.
This means that the combination phase
in diskmining is a simple union.

To illustrate this decomposition,
let $\db$ contain the transactions
$\{a, b, d\}, \{b, c, d\}, \{a, c\}$ and $\{a, b\}$.
Suppose the minimum support is 50\%, 
then $\db_{cd}=\{\{a, b, d\}, \{b, c, d\}, \{a, c\}\}$,
$\db_{ab} =\{ \{a, b\}, \{b\}, \{a\}, \{a, b\}\}$.
From $\db_{cd}$, we get all frequent itemsets 
$\{d\}, \{b,d\}$, and $\{c\}$.
Note though $\{a\}$ and $\{b\}$ are also frequent in $\db_{cd}$,
they're not listed since they contain neither $c$ nor $d$.
They will be listed in the frequent itemsets of $\db_{ab}$,
which are $\{a\}, \{b\}$, and $\{a,b\}$. 

To analyze the recurrence and required disk I/O's of the general 
divide-and-conquer algorithm
when the decomposition strategy is projection,
let us suppose that:

\begin{small}
\begin{list}{-}{}

\item
The original database size is $D$ bytes.

\item
The data structure is an FP-tree.

\item
The FP-tree constructed from original database \db~is $T$, 
and its size is $|T|$ bytes.

\item
If a conditional FP-tree $T'$ is constructed from 
an FP-tree $T$, then $|T'|\leq c\cdot |T|$,
for some constant $c<1$.

\item
The main memory mining method is the {\em FP-growth} 
method \cite{HPY00, HPYM04}.
Two database scans are needed for constructing an FP-tree
from a database.

\item
The block size is $B$ bytes.

\item 
The main memory available for the FP-tree is $M$ bytes

\end{list}
\end{small}

In the first line of the algorithm in \xf{bdaqalgo},
if $T$ can not fit in memory,
then projected databases will be generated.
We assumed that
the size of the FP-tree for a projected database
is  $c\cdot|T|$. 
If $c\cdot |T| \leq M$, function 
{\em mainmine} can be called for the projected database,
otherwise, the decomposition goes on.
At pass $m$, the size of the FP-tree constructed from 
a projected database is $c^m\cdot |T|$.
Thus, the number of passes needed by the 
divide-and-conquer projection algorithm is   
$1+\lceil\log_cM/T\rceil$.
Based on our experience and the analysis in \cite{HPY00, HPYM04},
we can say that for all practical purposes
the number of passes will be at most two.
For example, Let $D = 100$ Giga and $T = 10$ Giga, 
$M = 1$ Giga, $c = 10\%$. 
Then the number of passes is
$1+\lceil\log_{0.1}2^{30}/(10\times 2^{30})\rceil$ = 2. 
In five passes we can handle databases up to 100 Terabytes.
Namely, we get
$1+\lceil\log_{0.1}2^{30}/(10\times 2^{40})\rceil$ = 5.

Assume that there are two passes,
and that the sum of the sizes of all projected
databases is $D'$.
There are two database scans for \db, 
one for finding all frequent single items,
one for decomposition.
Two scans need $2\times D/B$ disk I/O's. 
The projected databases have to be written to the disks first,
then later each scanned twice for building the FP-tree.
This step needs  $3\times D'/B$ disk I/O's.
Thus, the total disk number of 
disk I/O's for the general divide-and-conquer 
projection algorithm
is at least 
\negvs
\begin{eqnarray}
2\cdot D/B + 3\cdot D'/B.
\end{eqnarray}
Obviously, 
the smaller $D'$, the better the performance.

One of the simplest projection strategies
is to project the database on each frequent item,
which we call
{\em naive projection}.
First we need some formal definitions.

\begin{defn}
{\rm
Let $I$ be a set of items.
By $I^*$ we will denote {\em strings} over $I$,
such that each symbol occurs at most once in the string.
If $\alpha$, $\beta$ are strings, and $i_j$ an item,
then 
$\alpha.\beta$ denotes the concatenation of the
string $\alpha$ with the string $\beta$. 

For a string $\alpha$, we shall denote
by $\{\alpha\}$, the {\em set} of items occurring in it.

Let $\db$ be an $I$-database.
Then ${\mit freqstring}(\db)$
is the string over
$I$, such that each frequent item in $\db$ occurs
in it exactly once, and the items are in decreasing
order of frequency in $\db$.
\hspace*{\fill}${\qed}$
}
\end{defn}

As an example, consider the  $\{a,b,c,d\}$-database
$\db = \{\{a,b,c\}, \{a,b,c,d\}, \{a,c\}\}$.
If the minimum support is 60\%, then
${\mit freqstring}(\db) = acb$.
Note that $\{acb\} = \{a,c,b\}$.

\begin{defn}
{\rm 
Let $\db$ 
be an $I$-database, and let 
${\mit freqstring}(\db) 
= i_1i_2\cdots i_k$.
For $j\in\{1,\ldots,k\}$ we define
$\db_{i_j} =
\{\tau\cap\{i_1,\ldots,i_j\} : i_j\in\tau,\tau\in\db\}.$

Let $\alpha\in I^*$.
We define $\db_{\alpha}$ inductively:
$\db_{\epsilon} = \db$, and
let ${\mit freqstring}(\db_{\alpha}) 
= i_1i_2\cdots i_k$. Then,
for $j\in\{1,\ldots,k\}$, 
$\db_{\alpha.i_j} =
\{\tau\cap\{i_1,\ldots,i_j\} : i_j\in\tau,\tau\in\db_{\alpha}\}.$
\hspace*{\fill}${\qed}$
}
\end{defn}

Obviously,  
$\db_{\alpha.i_j}$ is an $\{i_1,\ldots,i_j\}$-database.
The decomposition of $\db_{\alpha}$ into
$\db_{\alpha.i_1}$, \ldots, $\db_{\alpha.i_k}$
is called the {\em naive projection}.

\begin{defn}
{\rm 
Let $\alpha\in I^*$, $i_j\in I$, and let
$\db_{\alpha.i_j}$ be an $I$-database.
Then ${\mit freqsets}(\xi,\db_{\alpha.i_j})$ denotes the subsets
of $I$ 
that contain $i_j$ and are frequent in $\db_{\alpha.i_j}$ 
when the  minimum support is $\xi$.
Usually, we shall abstract $\xi$ away, and write
just  ${\mit freqsets}(\db_{\alpha.i_j})$
\hspace*{\fill}${\qed}$
}
\end{defn}

\begin{lemma}

Let $\db_{\alpha}$ be an $I$-database, and
${\mit freqstring}(\db_{\alpha}) = i_1i_2\cdots i_k$.
Then
$${\mit freqsets}(\db_{\alpha}) =
\bigcup_{j\in\{1,\ldots,k\}}{\mit freqsets}(\db_{\alpha.i_j})$$

\end{lemma}

\noindent
{\bf Proof}. 
($\subseteq$-{\em direction}). 
Let $S\in {\mit freqsets}(\db_{\alpha})$,
and suppose $i_n$ is the item in $S$ that is least frequent in
$\db_{\alpha}$.
Since $\db_{\alpha.i_n}$ is an $\{i_1,\ldots,i_n\}$-database,
and transactions in $\db_{\alpha}$ that contain item $i_j$ 
are all in $\db_{\alpha.i_j}$,
if $S$ is frequent in $\db_{\alpha}$, 
then $S$ must be frequent in $\db_{\alpha.i_j}$.

\noindent
($\supseteq$-{\em direction}). 
For any frequent itemset 
$S \in freqsets(\db_{\alpha.i_j})$,
according to the definition,
the 
support of any itemset in $\db_{\alpha.i_j}$ is not greater than
the support of it in $\db_{\alpha}$.
Therefore, $S$ must be frequent in $\db_{\alpha}$.
\hspace*{\fill}${\qed}$

\medskip

\xf{hansalgo} gives a divide-and-conquer algorithm
that uses naive projection.
A transaction $\tau$ in $\db_{\alpha}$ will be partly inserted into 
$\db_{\alpha.i_j}$ if and only if $\tau$ contains $i_j$.
The parallel projection algorithm introduced in
\cite{HPYM04} 
is an algorithm of this kind.

\begin{figure}[h]
{\bf Procedure} {\em naivediskmine}($\db_{\alpha},M$)

\smallskip

{\bf if} $|\db_{\alpha}|\leq M$ {\bf then} 
{\bf return} {\em mainmine($\;\db_{\alpha}$)} 

{\bf else} let ${\mit freqstring}(\db_{\alpha}) = i_1i_2\cdots i_n$

{\hskip 18pt} {\bf return}  {\em naivediskmine}$(\db_{\alpha.i_1},M)\;\cup$ 

{\hskip 146pt}  $\ldots\;\cup$ 

{\hskip 56pt}{\em naivediskmine}$(\db_{\alpha.i_n},M)$.

\caption{{\small 
A simple divide-and-conquer algorithm for 
mining frequent itemsets from disk 
}}
\label{hansalgo}
\end{figure}

Let's analyze the disk I/O's of the algorithm
in \xf{hansalgo}.
As before, we assume that there are two passes,
that the data structure is an FP-tree,
and that the main memory mining method is
{\em FP-growth}. 
If in $\db_{\epsilon}$, each transaction contains on the average $n$
frequent items, 
each transaction will be written to $n$ projected databases.
Thus the total length of the associated transactions in
the projected databases is
$n+(n-1)+\cdots+1 = n(n+1)/2$,
the total size of all projected databases is
$(n+1)/2\cdot D\approx n/2\cdot D$.

There are two database scans for $\db_{\epsilon}$,
one for finding all frequent single items,
and one for decomposition.
Two scans need $2\cdot D/B$ disk I/O's.
The projected databases have to be written to the disks first,
then later scanned twice each for building an FP-tree.
This step needs at least $3\cdot n/2\times D/B$.
Thus, the total disk I/O's for the divide-and-conquer
algorithm with naive projection 
is 
\negvs
\begin{eqnarray}
2 \cdot D/B 
+ 
n \cdot 3/2 \cdot D/B
\end{eqnarray}

The recurrence structure of algorithm
{\em naivediskmine} 
is shown in \xf{naivetree}.
The reader should ignore
nodes in 
the shaded area
at this point, they
represent processing
in main memory.

\begin{figure}[h]
\centerline{\psfig{figure=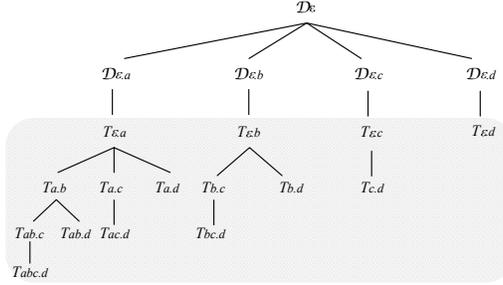,height=1.5in}}
\caption{\small Recurrence structure of Naive Projection}
\label{naivetree}
\end{figure}

In a typical application $n$, the average number
of frequent items could be hundreds, or thousands.
It therefore makes sense to devise a smarter
projection strategy.
Before we go further, we introduce
some definitions and a lemma.

\begin{defn}\label{four}
{\rm
Let $\db_{\alpha}$ be an $I$-database, and let
${\mit freqstring}(\db_{\alpha}) 
= \beta_1.\beta_2. \cdots .\beta_k$,
where each $\beta_j$ is a string in $I^*$.
We call $\beta_1.\beta_2. \cdots .\beta_k$
a {\em grouping} of 
${\mit freqstring}(\db_{\alpha})$. 
For 
$j\in\{1,\ldots,n\}$,
we now define
$\db_{\alpha.\beta_j} =
\{\tau\cap\{\beta_1,\ldots,\beta_j\} : \tau\in\db_{\alpha},
\tau\cap\beta_j\neq\emptyset
\}.$

In $\db_{\alpha.\beta_j}$,
items in $\{\beta_j\}$ are called {\em master items},
items in $\{\beta_1,\ldots,\beta_{j-1}\}$ are called {\em slave items}.
\hspace*{\fill}${\qed}$
}
\end{defn}

For example,
if ${\mit freqstring}(\db_{\alpha}) = abcde$,
$\beta_1 = abc$, $\beta_2 = de$ gives
the grouping $abc.de$ of $abcde$.

\begin{defn}
{\rm
Let $\{\alpha,\beta\}\subset I^*$, and let
$\db_{\alpha.\beta}$ be an $I$-database.
Then $freqsets(\db_{\alpha.\beta})$ denotes the subsets
of $I$ 
that contain at least one item in $\{\beta\}$
and are frequent in $\db_{\alpha.\beta}$.
\hspace*{\fill}${\qed}$
}
\end{defn}

\begin{lemma}\label{goodway}
Let $\alpha\in I^*$, 
$\db_{\alpha}$ be an $I$-database, and
${\mit freqstring}(\db_{\alpha}) = \beta_1\beta_2\cdots \beta_k$.
Then
$$freqsets(\db_{\alpha}) =
\bigcup_{j\in\{1,\ldots,k\}}freqsets(\db_{\alpha.\beta_j})$$

\end{lemma}

\noindent
{\bf Proof.}
Straightforward from Lemma 1 and the definition 
of $\db_{\alpha.\beta}$.
\hspace*{\fill}${\qed}$

\medskip

Based on Lemma \ref{goodway},
we can obtain a more aggressive divide-and-conquer algorithm for 
mining from disks.
\xf{ouralgo} shows the algorithm {\em aggressivediskmine}.
Here,
${\mit freqstring}(\db_{\alpha})$
is decomposed into several substrings $\beta_j$,
each of which could have more than one item.
Each substring corresponds to a projected database.
A~transaction $\tau$ in $\db_{\alpha}$ will be partly inserted into
$\db_{\alpha.\beta_j}$ if and only if 
$\tau$ contains at least one item $a$
such that $a\in\{\beta_j\}$.
Since there will be fewer projected databases,
there will be less disk I/O's.
Compared with the algorithm in \xf{hansalgo},
we can expect that
a large amount of disk I/O will be saved by the algorithm
in \xf{ouralgo}.

\begin{figure}[h]
{\bf Procedure} {\em aggressivediskmine}($\db_{\alpha},M$)

\smallskip

{\bf if} $|\db_{\alpha}|\leq M$ {\bf then} 
   {\bf return} {\em mainmine($\;\db_{\alpha}$)} 

{\bf else} let ${\mit freqstring}(\db_{\alpha}) = 
\beta_1\beta_2\cdots \beta_k$

{\hskip 18pt}
{\bf return}  {\em aggressivediskmine}$(\db_{\alpha.\beta_1},M)\;\cup$ 

{\hskip 165pt} $\;\ldots\;\cup$ 

{\hskip 57pt}{\em aggressivediskmine}$(\db_{\alpha.\beta_k},M)$.

\caption{{\small 
A more aggressive divide-and-conquer algorithm for 
mining frequent itemsets from disk 
}}
\label{ouralgo}
\end{figure}

Let's analyze the recurrence and disk I/O's of the aggressive 
divide-and-conquer algorithm.
The number of
passes needed by the algorithm is still 
\mbox{$1+\lceil\log_cM/T\rceil \approx 2$},
since grouping items doesn't change the size of an FP-tree for
a projected database.
However, for disk I/O,
suppose in $\db_{\epsilon}$, 
each transaction contains on average $n$
frequent items,
and that we can group them into $k$
groups of equal size.
Then the $n$ items will be written to the projected databases
with total length $n/k+2\cdot n/k+ \ldots +k\cdot n/k = (k+1)/2\cdot n$.
Total size of all projected databases is
$(k+1)/2\cdot D \approx k/2\cdot D$.
The total disk I/O's for the aggressive divide-and-conquer
algorithm
is then 
\negvs
\begin{eqnarray}\label {formula}
2\cdot D/B 
+
k \cdot 3/2 \cdot D/B
\end{eqnarray}

The recurrence structure of algorithm
{\em aggressivediskmine} is shown
in \xf{recagg}. Compared to \xf{naivetree},
we can see that the part of the tree
that corresponds to decomposition
(the nonshaded part) is much smaller
in \xf{recagg}. Although the example is
very small, it exhibits the general structure
of the two trees.

\begin{figure}[h]
\centerline{\psfig{figure=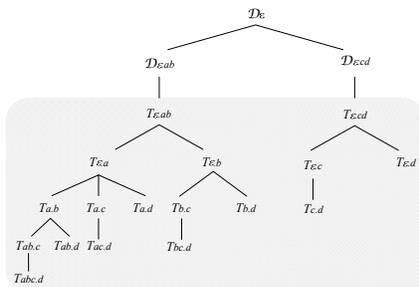,height=1.5in}}
\caption{\small Recurrence structure of Aggressive Projection}
\label{recagg}
\end{figure}

If $k\ll n$,
we can expect that the aggressive
divide and conquer algorithm will
significantly outperform the naive one.

\section {Algorithm Diskmine}
In this section
we give
the details of
our divide-and-conquer algorithm for mining frequent itemsets
from secondary memory.
We call the algorithm {\em Diskmine}.
In the algorithm, 
the FP-tree is used as data structure and 
the extension of {\em FP-growth} method, 
{\em FP-growth*} \cite{fimi03},
as method for mining frequent itemsets from an FP-tree.
Before introducing the algorithm, 
let's first recall the FP-tree and the {\em FP-growth* } method.

\subsection{The FP-tree and {\em FP-growth*} method}

The {\em FP-tree (Frequent Pattern tree)}
is a data structure used in 
the {\em FP-growth} method by Han {\em et al.\ } \cite{HPY00}.
It is a compact representation
of all relevant
frequency information
in a database.
The nodes of the FP-tree stores an item name, item count,
and a link.
Every branch of the FP-tree represents a frequent itemset,
and the nodes along the branches are
stored in decreasing order of the frequency
of the corresponding items, with leaves representing
the least frequent items.
Compression is achieved by
building the tree in such a way that
overlapping itemsets
share prefixes of the
corresponding branches.

The FP-tree has
a {\em header table} associated with it.
Single items and their counts are stored in
the header table in
decreasing order of their frequency.
The entry for an item also contains the head
of a list that links all the
nodes of the item
in the FP-tree.

The FP-growth method needs two database scans
when mining all frequent itemsets.
The first scan counts the number of occurrences
of each item.
The second scan constructs the initial FP-tree,
which contains all frequency information of the original dataset.
Mining the database then becomes mining the FP-tree.

The {\em FP-growth} method relies on the following
principle: if $X$ and $Y$ are two itemsets,
the count of itemset $X\cup Y$ in the database
is exactly that of $Y$ in the restriction of the database to
those transactions containing $X$.
This restriction of the database is
called
the {\em conditional pattern base} of $X$,
and the FP-tree constructed from the conditional pattern base
is called $X$'s {\em conditional FP-tree},
which we denote by $T_X$.
We can view the FP-tree constructed from the initial database
as $T_{\emptyset}$,
the conditional FP-tree for $\emptyset$.
Note that for
any itemset $Y$ that is frequent
in the conditional pattern base of $X$,
the set
$X\cup Y$ is a frequent itemset for the original database.\footnote{In
keeping with the notation introduced so far, we shall
in the sequel write $T_{\alpha}$ when we mean the 
FP-tree $T_{\{\alpha\}}$. Similarly we shall write
$T_{\alpha.i}$ instead of $T_{\{\alpha\}\cup\{i\}}$.}

The recursive structure of FPgrowth can be seen from
the shaded area in \xf{naivetree}.
In the figure, we will enter the main memory phase
for instance for the conditional database $\db_a$.
Then FP-growth first constructs the
FP-tree $T_a$ from $\db_a$.
The tree rooted at $T_a$
shows the recursive structure of FP-growth,
assuming for simplicity that the
relative frequency remains the same in
all conditional pattern bases.

In \cite {fimi03}, we extend the FP-growth method into the
{\em FP-growth*} method by using an {\em array technique} 
and other optimizations. 
The experimental results in the paper
and those done by the FIMI-organizers show
that the FP-growth* method outperforms the {\em FP-growth} method
especially when the database is big or sparse
\cite{fimi03,ZB03}.

\subsubsection* {The array technique} \label{arraytech}

In the original FP-growth method \cite{HPY00}, 
to construct an FP-tree from a database $\db$, 
two database scan are required.
The first scan gets all frequent items,
the second constructs the FP-tree.
And later,
for each item $a$ in
the header of a conditional FP-tree $T_{\alpha}$,
two traversals of $T_{\alpha}$ are needed for constructing
the new conditional FP-tree $T_{\alpha.i}$.
The first traversal finds all frequent items in the
conditional pattern base of $\alpha.i$,
and initializes the FP-tree  $T_{\alpha.i}$
by constructing its header table.
The second traversal constructs the new tree
$T_{\alpha.i}$.

In the boosted {\em FP-growth*} method \cite{fimi03},
a simple data structure, an array,
is introduced to omit the first scan of $T_{\alpha}$.
This is achieved 
by constructing an array
$A_{\alpha}$ while building $T_{\alpha}$.
More precisely,
in the second scan of the original database we 
construct $T_{\epsilon}$, and an array $A_{\epsilon}$.
The array will store the counts of
all 2-itemsets, each cell $[j,k]$
in the array is a counter of the 2-itemset $\{i_j,i_k\}$.
All cells in the array are initialized to 0.
When an itemset is inserted into $T_{\epsilon}$,
the associated cells in $A_{\epsilon}$ are updated.
After the second scan,
the array $A_{\epsilon}$ contains the counts of
all pairs of items frequent in $\db_{\epsilon}$.

Next, the {\em FP-growth*} method is recursively called
to mine frequent itemsets for each item in header table
of $T_{\epsilon}$.
However, now for each item $i$,
instead of traversing $T_{\epsilon}$ along
the linked list starting at $i$ to get
all frequent items in $i$'s conditional pattern base,
$A_{\epsilon}[i,*]$ gives all frequent items for $i$.
Therefore, for each item $i$ in $T_{\epsilon}$
the array $A_{\epsilon}$ makes
the first traversal of $T_{\epsilon}$ unnecessary,
and $T_{\epsilon.i}$ can be
initialized directly from $A_{\epsilon}$.

For the same reason, from a conditional FP-tree $T_{\alpha}$,
when we construct a new conditional
FP-tree for $\alpha.i$, for an item $i$,
a new array $A_{\alpha.i}$ is calculated.
During the construction of the
new FP-tree $T_{\alpha.i}$,
the array $A_{\alpha.i}$
is filled.
The construction of arrays and FP-trees continues
until the {\em FP-growth} method terminates.

Note that if for a database, 
if we have the array that stores the count of all pairs of 
frequent items,
then only one database scan is needed 
to construct an FP-tree from the database.

\subsection{Divide-and-conquer by aggressive projection}

The algorithm {\em Diskmine} is shown in \xf{appa}. In the algorithm,
$\db_{\alpha}$ is the original database or a projected database,
and $M$ is the maximal size of main memory that can be used by {\em Diskmine}. 

\begin{figure}[h]
{\bf Procedure} {\em Diskmine}$(\db_{\alpha}, M)$

\smallskip

scan $\db_{\alpha}$ and compute {\it freqstring}$(\db_{\alpha})$

call ${\mit trialmainmine(\db_{\alpha}, M)}$ 

{\bf if} ${\mit trialmainmine(\db_{\alpha}, M)}$ aborted {\bf then}

{\hskip 12pt}compute a grouping $\beta_1\beta_2\cdots \beta_k$
   of ${\mit freqstring}(\db_{\alpha})$.
 
{\hskip 12pt}Decompose $\db_{\alpha}$ into
$\db_{\alpha.\beta_1},\ldots, \db_{\alpha.\beta_k}$

{\hskip 12pt}{\bf for} j = 1 {\bf to} k {\bf do begin} 

{\hskip 24pt}{\bf if} $\{\beta_j\}$ is a singleton {\bf then}  

{\hskip 36pt}${\mit Diskmine}(\db_{\alpha.\beta_j},M)$

{\hskip 24pt}{\bf else}  

{\hskip 36pt}${\mit mainmine}(\db_{\alpha.\beta_j})$

{\hskip 12pt}{\bf end}

{\bf else return} {\em freqsets}$(\db_{\alpha})$
\caption{{\small Algorithm Diskmine}}
\label{appa}
\end{figure}

{\em Diskmine} uses the FP-tree as 
data structure and {\em FP-growth*} \cite{fimi03}
as main memory 
mining
algorithm. 
Since the FP-tree encodes all frequency information
of the database, 
we can shift into main memory mining
as soon as the FP-tree fits
into main memory.

Since an FP-tree usually is a significant
compression of the database, our {\em Diskmine}
algorithm begins optimistically, by calling {\em trialmainmine},
which starts scanning the database and constructing the FP-tree.
If the tree can be successfully completed and stored in main memory,
we have reached the bottom level of the recursion,
and can obtain 
the frequent itemsets of the database
by running
{\em FP-growth*} on the FP-tree in main memory.

\begin{figure}[h]
{\bf Procedure} {\em trialmainmine}$(\db_{\alpha}, M)$

start scanning $\db_{\alpha}$ and building the FP-tree 
   
   {\hskip 12 pt}$T_{\alpha}$ in main memory.

{\bf if} $|T_{\alpha}|$  exceeds  $M$ {\bf then}

{\hskip 12pt}{\bf return} the incomplete $T_{\alpha}$ 

{\bf else} 

{\hskip 12pt}call {\em FP-growth*}$\,(T_{\alpha})$ and {\bf return}
    {\em freqsets}$(\db_{\alpha})$.

\caption{{\small Trial main memory mining algorithm}}
\label{trial}
\end{figure}

If, at any time during {\em trialmainmine}
we run out of main memory, we abort and
return the partially constructed FP-tree,
and a pointer to where we stopped scanning the database.
We then resume processing {\em Diskmine}$(\db_{\alpha},M)$
by computing a grouping 
$\beta_1,\ldots, \beta_k$ of 
{\em freqstring}$(\db_{\alpha})$, 
and then decomposing
$\db_{\alpha}$ into
$\db_{\alpha.\beta_1},\ldots,\db_{\alpha.\beta_k}$.
We recursively process 
each decomposed database
$\db_{\alpha.\beta_j}$.
During the first level of the recursion,
some groups $\beta_j$ will consist of a single
item only. 
If $\{\beta_j\}$ is a singleton,
we call {\em Diskmine}, otherwise
we call {\em mainmine} directly,
since we put several items in a group
only when we estimate that the corresponding
FP-tree will fit into main memory.

In computing the grouping
$\beta_1,\ldots, \beta_k$ 
we assume that transactions in a very large database
are evenly distributed, i.e., 
if an FP-tree is constructed from part of a database,
then this FP-tree represents the whole FP-tree for the whole database.
In other words,
if the size of the FP-tree is $n$ for $p\%$ of the database,
then the size of the FP-tree for whole database is $n/p \cdot 100$.
Most of the time, this gives an overestimation,
since an FP-tree increases fast only at the beginning stage,
when items are encountered for the first time and inserted
into the tree. In the later stages, the changes to the FP-tree
will be mostly counter updates.

\begin{figure}[h]
{\bf Procedure} {\em mainmine}$(\db_{\alpha.\beta})$

build a modified FP-tree $T_{\alpha.\beta}$ for $\db_{\alpha.\beta}$

{\bf for each} $i\in\{\beta\}$ {\bf do begin}

{\hskip 12pt} construct the FP-tree $T_{\alpha.i}$
                for $\db_{\alpha.i}$ from $T_{\alpha.\beta}$
   
{\hskip 12pt} call {\em FP-growth*}$\,(T_{\alpha.i})$ 
    and {\bf return}
    {\em freqsets}$(\db_{\alpha.i})$.

{\bf end}

\caption{{\small Main memory mining algorithm}}
\label{mainmine}
\end{figure}

Since we know that there is only one master item in the database 
(for $\db_\epsilon$, no master item at all),
an FP-tree is constructed without the master item.
In \xf{mainmine},
since $\db_{\alpha.\beta}$ is for multiple master items,
the 
FP-tree constructed from $\db_{\alpha.\beta}$ has to contain
those master items.
However, the item order is a problem for the FP-tree,
because we only want to mine all frequent itemsets
that contain master items.
To solve this problem,
we simply use the item order in the partial FP-tree
returned by the aborted 
{\em trialmainmine}$(\db_{\alpha})$.
This is what we mean by a ``modified FP-tree''
on the first line in the algorithm in \xf{mainmine}.

The entire recurrence structure of
{\em Diskmine} can be seen in \xf{recagg}.
Compared to the naive projection in \xf{naivetree}
we see that since the aggressive projection
uses main memory more effective,
the decomposition phase is shorter,
resulting in less I/O.

\begin{theorem}
Diskmine$(\db)$  returns freqsets$(\db)$.
\end{theorem}

\noindent
{\bf Proof}. 
The correctness of {\em Diskmine}
can be derived from the correctness of the 
{\em FP-growth*} method in \cite{fimi03}
and Lemma \ref{goodway} in \xs{diskmine}.
In {\em Diskmine},
each item acts as master item in exactly one projected database.
If a projected database is only for one master item $i_j$,
the result of  {\em FP-growth*} method or a recursive call of {\em Diskmine} 
will be $freqsets(\db_{i_j})$. 
If a projected database is for a set $\{\beta\}$ of master items,
it contains all frequency information associated with the master items.
Since in the {\em FP-growth*} method,
the order of the items in an FP-tree doesn't influence 
the correctness of the  {\em FP-growth*} method,
{\em mainmine} indeed returns only frequent itemsets that 
contain master item(s),
i.e.\ {\em mainmine} gives the 
exact value of $freqsets(\db_{\alpha.\beta})$.
According to Lemma \ref{goodway},
algorithm {\em Diskmine} then
correctly outputs all 
itemsets in frequent the original database.
\hspace*{\fill}${\qed}$

\subsection {Memory Management}\label{memory}

Given a database $\db_{\alpha}$, 
to successfully apply the {\em FP-growth*} method, 
the basic main memory requirement is that the size of the FP-tree
$T_{\alpha}$
constructed from $\db_{\alpha}$,
is less than the available amount $M$ of main memory.
In addition, we need space
for the  descendant conditional 
FP-trees that will be constructed during the recursive calls
of {\em FP-growth*}.

Suppose the main memory requirement 
for $T_{\alpha}$ plus its descendant FP-trees is $m$.
If $M < m$, but the difference $m-M$ is not very big, 
the {\em FP-growth*} method
could still be run because the operating 
system uses virtual memory.
However, there could be too many page swappings
which takes too much time and makes {\em FP-growth*} very slow.
Therefore, given $M$, for a very large database $\db_{\alpha}$,
we have to stop the construction of the FP-tree $T_{\alpha}$
and the execution of {\em FP-growth*} method before
all physical main memory is used up.

Another problem is that we will 
construct a large number  of FP-trees.
Since there can be 
millions of nodes in those FP-trees,
inserting and deleting nodes is time consuming.
 
In the implementation of the algorithm,
we use our own main memory management for 
allocating and deallocating nodes,
and calculating the main memory we have already used.
We assume that the main memory needed by an FP-tree is
proportional to the number of nodes in the FP-trees.
We also assume that the workspace needed for calling  
{\em FP-growth*(T)} method on an FP-tree is roughly 10\%
of the size of the FP-tree $T$.
Here, 10\% is a liberal assumption according to the
experimental result in \cite{HPY00}. 
Later in this section, a more accurate value will be given.
If the size of FP-tree is more than $0.9\cdot M$,
we conclude that $M$ is not big enough to store whole 
FP-tree $T_{\alpha}$.

Since all memory for nodes in an FP-tree is deallocated after a call 
of {\em FP-growth*} ends,
a chunk of memory is allocated for each FP-tree when we create the tree,
and the chunk size is changeable. 
After generating all frequent 
itemsets from the FP-tree, the chunk is discarded, 
and all nodes in the tree are deleted.
Thus we successfully avoid freeing nodes in FP-trees one by one,
which would take too much time.

\subsection{Applying the Array Technique}\label{array}

In {\em Diskmine}, 
the array technique is also be applied to save FP-tree traversals.
Furthermore, when projected databases are generated,
the array technique can save a great number of disk~I/O's.

Recall that in {\em trialmainmine},
if an FP-tree can not be accommodated in main memory,
the construction stops. 
Suppose now we decided to stop
scanning the database.
Then later, after generating all projected databases,
for a projected database with only one master item,
two database scans are required to construct an FP-tree for the master item.
The first scan gets all frequent items for the master item,
the second scan constructs the FP-tree.
For a projected database with several master items,
though the FP-tree constructed from the database
uses the modified item order
(the order from the header of the FP-tree  in
the previous level of the recursion),
to construct new FP-trees for the master items,
two FP-tree traversals are needed.
To avoid the extra scan,
in {\em Diskmine} we calculate an array for each FP-tree.
When constructing the FP-tree from $\db_{\alpha}$, 
if it is found that the tree can not fit in main memory,
the construction of the FP-tree $T_{\alpha}$ stops,
but the scan of the database $\db_{\alpha}$
continues so that we finish filling the cells of 
the array $A_{\alpha}$.
Here, some extra disk I/O's are spent,
but the payback will be that we 
save one database scan for each
projected database.
Furthermore, finishing the scanning
of $\db_{\alpha}$ 
doesn't require any more main memory, 
since the array $A_{\alpha}$
is already there.

From the array, for each projected database,
the count of each pair of master items and 
the count of each pair of master item and slave item
can be known.
As an example,
suppose a projected databases is only for one
master item $i_j$
and slave items $i_1, \ldots, i_{j-1}$.
To mine all frequent itemsets,
from the line for $i_j$ in the array,
accurate counts for 
$[i_j, i_{j-1}],
[i_j, i_{j-2}],
\ldots,
[i_j, i_1]$
can be easily found.
If there were no array
we would need an extra database scan.

With the array, we can also make a projected database 
drastically smaller.
In the definition of $\db_{\alpha.\beta_j}$,
we see that 
$\db_{\alpha.\beta_j}$ is an $\{\beta_1,\ldots,\beta_j\}$-database.
Actually, by checking the array $A_{\alpha}$,
if a slave item is found not frequently co-occurring 
with any master item in $\beta_j$,
it's useless to include the slave item in $\db_{\alpha.\beta_j}$,
because no frequent itemsets mined from $\db_{\alpha.\beta_j}$
will contain that slave item.
For same reason, 
if we also find that a master item $a$ is not frequent with any 
other master item or slave item, 
it will be not written to $\db_{\alpha.\beta_j}$, 
either. 
However, the frequent itemset $\alpha.a$ is outputted. 
Furthermore,
if from the array, we see that a  master item $a$ is
only frequent with one item (master or slave) $b$,
frequent itemsets $\alpha.a$ and $\alpha.a.b$
are outputted directly, 
and item $a$ will not appear in $\db_{\alpha.\beta_j}$.  
Therefore, by looking through the array, 
we find all slave items, 
such that they are not frequent with any master item in $\beta_j$,
and all master items, such that their number of frequent items in 
$\{\beta_1,\ldots,\beta_j\}$ is 0 or 1.
When generating $\db_{\alpha.\beta_j}$,
all those items are removed from the 
transactions we put in $\db_{\alpha.\beta_j}$.

\subsection{Statistics}

\begin{table*}[ht!]
\centering
\begin{tabular}
{|r|l|} \hline
$\tdb$&Number of transactions in $\db_{\alpha}$\\
\hline
$A_{\alpha}[j,k]$&Count of frequent item pair $\{i_j, i_k\}$ 
in $\db_{\alpha}$\\
\hline
$\tT$&Number of transactions used for constructing  $T_{\alpha}$\\
\hline
$\nT$&Number of nodes in $T_{\alpha}$\\
\hline
$\njT$&Number of nodes in  $T_{\alpha}$ if we retain 
only nodes for items $i_1, \ldots, i_j$\\
\hline
$\mjT$&Number of nodes in  
$T$, 
where a  node $P$ for item $i_k$ is counted if\\
&it satisfies the following conditions: 1) $P$ is in a branch that contains $i_j$\\
&2) $i_k \in \{i_1, \ldots, i_j\}$ 3) $A_{\alpha}[j,k] > \xi$\\
\hline

\end{tabular}
\caption{Statistics Information}
\label{stat}
\end{table*}

Algorithm
{\em Diskmine} collects some statistics on the
partial FP-tree $T_{\alpha}$ 
and the rest of database $\db_{\alpha}$,
for the purpose of
grouping items together.
\xt{stat} shows the statistics information.
In the table, 
$\db_{\alpha}$ is the original database or the current projected database,
and {\em freqstring}($\db_{\alpha}$)=
$i_1\ldots i_j\ldots i_k \ldots i_n$.
The partial FP-tree is $T_\alpha$ 
and $\xi$ is the 
absolute value of the minimum support.

In the table, 
the array discussed in \xs{array}
is also listed as statistics.
Values for the cells of 
the array are accumulated during the construction of
the partial $T_{\alpha}$.
If {\em trialmainmine} is aborted, the rest
of the statistics 
is collected by scanning the
remaining part of $\db_{\alpha}$.
Values in  
$\njT$ 
can also be obtained
during the construction of $T_{\alpha}$.
Here
$\njT$ 
records the size of the FP-tree after
$T_{\alpha}$ is trimmed and only contains items $i_1, \ldots, i_j$.
Notice that
$\nT$ 
is equal to 
$\nu[n](T_{\alpha})$.
This is  also the size of a tree that can fit in main memory.
The value for  
$\mjT$
can be obtained
by traversing $T_{\alpha}$ once,
it gives the size of the FP-tree $T_{\alpha.i_j}$.

It might seem that 
collecting all this statistics
is a large overhead,
however, 
since all work is done in main memory,
it doesn't take much time.
And the time saved for disk I/O's 
is far more than the time spent on gathering statistics.

\subsection{Grouping items}

In \xf{appa},
the fourth line computes a grouping $\beta_1\beta_2\cdots \beta_k$
of ${\mit freqstring}(\db_{\alpha})$.
Each string $\beta$ 
corresponds to a group and each $\beta$ consists of at least one item.
For each $\beta$, 
a new projected database $\db_{\alpha.\beta}$
will be computed from $\db_{\alpha}$, 
then written to disk and read from disk later.
Therefore,
the more groups, 
the more disk I/O's. 
In other words,
there should be as many items in each
$\beta$ as possible. 
To group items,
two questions have to be answered.
\begin{enumerate}
\item If $\beta$ currently only has one item $i_j$, 
after projection, is the main memory big enough for
accommodating $T_{\alpha.i_j}$ constructed from 
$\db_{\alpha.i_j}$
and running the {\em FP-growth*} method on $T_{\alpha.i_j}$?
\item If more items are put in $\beta$,
after projection, is the main memory big enough for
accommodating $T_{\alpha.\beta}$ constructed from $\db_{\alpha.\beta}$
and running {\em FP-growth*} on $T_{\alpha.\beta}$ only
for items in $\beta$?
\end{enumerate}

Answering the first question is pretty easy,
since for each item $i_j$, 
the number
$\mjT$
gives the size of an FP-tree if the tree
is constructed from the partial FP-tree $T_{\alpha}$.
Therefore 
$\mjT$
can be used to estimate the 
size of FP-tree $T_{\alpha.i_j}$.
By the assumption that
the transactions in $\db_{\alpha}$ are evenly
distributed and that
the partial $T_{\alpha}$ 
represents
the whole FP-tree for $\db_{\alpha}$,
the estimated size of FP-tree $T_{\alpha.i_j}$
is 
$\mjT\cdot \tdb/\tT$.

Before answering the second question,
we introduce the {\em cut point}
from which the first group can be easily found.

\medskip

\noindent
{\bf Finding the cut point.} 
Recall the order that {\em FP-growth*} uses in mining frequent itemsets.
Starting from the least frequent item $i_n$,
all frequent itemsets that contains $i_n$ are mined first.
Then the process is repeated for
$i_{n-1}$, and so on.
Notice that when mining frequent itemsets for $i_k$,
all frequency information about $i_{k+1},\ldots,i_n$ is useless.
Thus, though a complete FP-tree $T_\alpha$ constructed from $\db_\alpha$
could not fit in main memory,
we can find many $k$'s such that the 
trimmed FP-tree containing only 
nodes for items $i_k, \ldots, i_1$
will fit into main memory.
All frequent itemsets for  $i_k, \ldots, i_1$
can be then mined from one trimmed tree.
We call the biggest of such $k$'s the {\em cut point}.
At this point, main memory is big enough 
for storing the FP-tree
containing only $i_k, \ldots, i_1$, 
and there is also enough main memory for running
{\em FP-growth*} on the tree.
Obviously, if the cut point $k$ can be found, 
items  $i_k, \ldots, i_1$ can be grouped together. 
Only one projected database is needed for $i_k, \ldots, i_1$.

There are two ways to estimate the cut point.
One way is to get cut point from the value of 
$\tdb$
and 
$\tT$
in \xt{stat}.
\xf{divi} illustrates the intuition behind the cut point.
In the figure,
since the partial FP-tree for 
$\tT$
of 
$\tdb$
transactions can be 
accommodate in main memory,
we can expect that the FP-tree containing  $i_k, \ldots, i_1$,
where 
$k=\lfloor n \cdot \tT /\tdb \rfloor$,
also will fit in main memory.

\begin{figure}[h]
\centerline{\psfig{figure=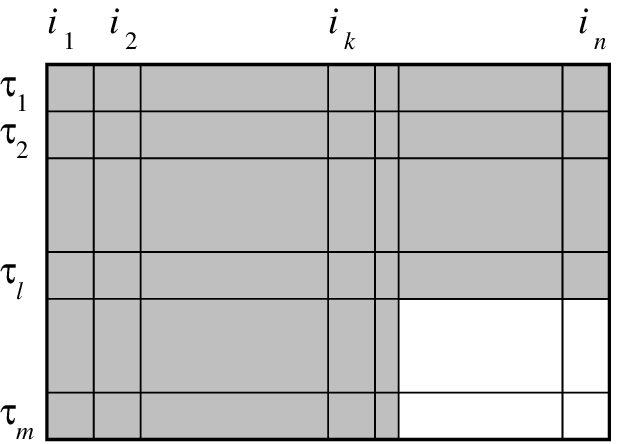,height=1.25in}}
\caption{Cut Point. Here
$l=\tT$, and $m=\tdb$}
\label{divi}
\end{figure}

The above method 
works well
for many databases, 
especially for those databases whose corresponding 
FP-trees have plenty of sharing of prefixes for items
from $i_1$ to the cut point.
However, 
if the FP-tree constructed from a database
doesn't share prefixes that much,
the estimation could fail, 
since now the FP-tree 
for items from $i_1$ to the cut point
could be too big.
Thus, 
we have to consider another method.
In \xt{stat},
$\njT$
records the size of the FP-tree after
the partial FP-tree $T_\alpha$ is trimmed and only 
contains items $i_1, \ldots, i_j$.
Based on 
$\njT$
the number of nodes 
in the complete FP-tree
for item $i_j$ 
can be estimated as  
$\njT \cdot \tdb/\tT$.
Now, finding the cut point becomes finding the biggest $k$ such that
$\nu[k](T_{\alpha}) \cdot \tdb/\tT \leq \nT$, 
and
$\nu[k+1](T_{\alpha}) \cdot 
\tdb/\tT > \nT$.

Sometimes the above estimation only guarantees 
that the main memory is big enough for 
the FP-tree which contains all items between $i_1$ and the cut point, 
while it doesn't guarantee 
that the descendant trees from that FP-tree can fit in main memory. 
This is because the estimation doesn't consider the 
size of descendant trees correctly 
(in \xs{memory}, we assumed that the size of a conditional tree is 10\%
of its nearest ancestor tree).
Actually, from 
$\mjT$
we can get a more accurate estimation of the size of the 
biggest descendant tree.
To find the cut point,
we need to find the biggest $k$,
such that 
$(\nu[k](T_{\alpha}) + 
\mjT)\cdot 
\tdb/\tT \leq \nT$, and
$(\nu[k+1](T_{\alpha}) + 
\mu[m](T_{\alpha})) 
> \nT$,
where 
$j\leq k$, 
$\mjT = {\mit max}_{j\in\{1,\ldots,k\}}\mjT$,
and 
$m\leq k+1$, 
$\mu[m](T_{\alpha}) = {\mit max}_{m\in\{1,\ldots,k+1\}}\mu[m](T_{\alpha})$.

\medskip

\noindent
{\bf Grouping the rest of the items.}
Now we answer the second question, how to put more items into a group?
Here we still need 
$\mjT$.
Starting with  
\mbox{$\mu[{\mit cutpoint}+1](T_{\alpha})$},
we test if 
$\mu[{\mit cutpoint}+1](T_{\alpha})\cdot
\tdb/\tT > \nT$.
If not, we put next item {\mit cutpoint}+2 
into the group,
and test if 
\mbox{$(\mu[{\mit cutpoint}+1](T_{\alpha}) +
\mu[{\mit cutpoint}+2](T_{\alpha}) 
)$}
$\cdot \tdb/\tT > \nT$.
We repeatedly put next item in 
${\mit freqstring}(\db)$ into the group
until we reach an item $i_j$,
such that
$$\displaystyle\sum_{m={\mit cutpoint}+1}^{j}
\mu[m](T_{\alpha})\cdot 
\tdb/\tT > \nT.$$
Then starting from $i_j$, we put items into next group,
until all items find its group.

Why can we group items together?
This is because
even if we construct 
$T_{\alpha.i_j}, \ldots, T_{\alpha.i_k}$
from the projected databases 
$\db_{\alpha.\beta_{i_j}}, \ldots, \db_{\alpha.\beta_{i_k}}$
and put all of them into main memory, 
the main memory is big enough according to the grouping condition.
At this stage, $T_{\alpha.i_j}, \ldots, T_{\alpha.i_k}$
all can be constructed by scanning $\db_\alpha$ once.
Then we mine frequent itemsets from the FP-trees.
However, we can do better.
Obviously $T_{\alpha.i_j}, \ldots, T_{\alpha.i_k}$ overlap a lot,
and the total size of the trees is
definitely greater than the size of $T_{\alpha.\beta}$.
It also means that we can put more items into 
each $\beta$,
only if the size of $T_{\alpha.\beta}$ 
is estimated to fit in main memory.
To estimate the size of $T_{\alpha.\beta}$, part of 
$T_{\alpha}$ 
has to be traversed by following the links for 
the master items in $T_{\alpha}$.

\subsection {Database projection}
After all items have found their groups, 
the original database will be projected to small databases according to  
Definition \ref{four}.
To save disk I/O's, three techniques can be used:
\begin {enumerate}
\item
In a group $\beta$, if the number of master items is greater than 
half of the number of frequent items 
(this often happens in the group that contains cut point),
then $\db_{\alpha.\beta}$ is not necessary 
computed. 
To mine all frequent itemsets,
$T_{\alpha.\beta}$ can be directly constructed from $\db_{\alpha}$ 
by reading it once. 
This is because  $\db_{\alpha.\beta}$ 
is not much smaller than $\db_{\alpha}$,
while the disk I/O's for reading from $\db_{\alpha}$ once 
is less than the disk I/O's for writing and reading 
$\db_{\alpha.\beta}$ once.  

\item
Since 
the partial tree $T_{\alpha}$ 
now in main memory, 
records all frequency 
information of those transactions that have
been read so far, 
when computing projected databases,
the frequency information of those transactions 
can be gotten from $T_{\alpha}$.
Thus
disk I/O's are only spent on reading from those transactions
that did not contribute to $T_{\alpha}$.

\item
As discussed in \xs{array},
by using the array technique,
in group $\beta_j$, we find all slave items,
such that they are not frequent with any master item in $\beta_j$,
and all master items, such that their number of frequent 
items in $\{\beta_1,\ldots,\beta_j\}$ is 0 or 1.
When computing $\db_{\alpha.\beta_j}$,
all those items are removed from new transactions in $\db_{\alpha.\beta_j}$.
\end{enumerate}

\subsection {The disk I/O's}
Let's re-count the disk I/O's used in {\em Diskmine}.
From the first scan we get all frequent items in $\db_{\epsilon}$,
which needs $D/B$ disk I/O's.
In the second scan we construct a partial FP-tree $T_{\epsilon}$, 
then continue scanning the rest database for statistics,
which needs another $D/B$ disk I/O's.
Suppose then that $k$ projected databases have to be computed.
According to \xs{diskmine},
the total size of the projected databases is 
approximately $k/2 \cdot D$.
For computing the projected databases,
the frequency information in $T_{\epsilon}$ is reused,
so only part of $\db_{\epsilon}$ is read. 
We assume on average half of $\db_{\epsilon}$ is read at this stage, 
which means $1/2\cdot D/B$ disk I/O's.
Writing and later reading $k$ projected databases
will take $2\cdot k/2\cdot D/B = k\cdot D/B$ disk I/O's.
Suppose all frequent itemsets can be mined from the projected databases
without going to the third level.
Then the total disk I/O's is
\negvs
\begin{eqnarray}
3/2 \cdot D/B 
+
k\cdot D/B
\end{eqnarray}

Compared with formula \ref{formula}, 
{\em Diskmine} saves at least 
$k/2 \cdot D/B$
disk I/O's,
thanks to the various techniques used in the algorithm.

\section {Experimental Evaluation and Performance Study}

In this section, we present the results from
a performance comparison of 
{\em Diskmine} with the {\em Parallel Projection Algorithm} in 
\cite{HPYM04} and the {\em Partitioning Algorithm} introduced 
in \cite{SON95}. 
The scalability of {\em Diskmine} is also analyzed,
and the accurateness of our memory size
estimations are validated.

As mentioned in \xs{diskmine},
the Parallel Projection Algorithm is a naive divide-and-conquer
algorithm, 
since for each item a projected database is created.
For performance comparison, 
we implemented Parallel Projection Algorithm, 
by using {\em FP-growth} as main memory method,
as introduced in \cite{HPYM04}.
The
Partitioning Algorithm is also a divide-and-conquer algorithm.
We implemented 
the partitioning algorithm by using the Apriori implementation
\cite{gap}.
We chose this implementation, since
it was well written and easy to adapt
for our purposes.

We ran the three algorithms on 
both synthetic datasets and real datasets.
Some synthetic datasets have millions of transactions,
and the size of the datasets ranges from several megabytes to 
several  hundreds gigabytes. 
Without loss of generality,
only the results for some synthetic datasets and a real dataset
are shown here.

All experiments were performed on a 2.0Ghz Pentium 4 with
256 MB of memory under Windows XP.
For {\em Diskmine} and the Parallel Projection Algorithm,
the size of the main memory is given as an input.
For the Partitioning Algorithm, 
since it only has two database scans and each main-memory-sized partition
and all data structures for Apriori 
are stored into main memory, 
the size of main memory is not controlled,
and only the running time is recorded.

We first compared the performance of three algorithms on synthetic dataset.
Dataset {\em T100I20D100K} was generated from the
application of \cite{syns}.
The dataset has 100,000 transactions and 1000 items,
and occupies about 40 megabytes of memory.
The average transaction length is 100,
and the average pattern length is 20.
The dataset is very sparse and FP-tree constructed from the dataset
is bushy. 
For Apriori, 
a large number of candidate frequent itemsets 
will be generated from the dataset. 
When running the algorithms, the main memory size
was given as 128 megabytes.
\xf{SynReal}(a) shows the experimental result.
In the figure, ``Naive Algorithm'' 
represents the Parallel Projection Algorithm,
and 
``Aggressive Algorithm'' represents the {\em Diskmine} algorithm.

\begin{figure}[h]
    \begin{minipage}[t]{2in}
       \centerline{\psfig{figure=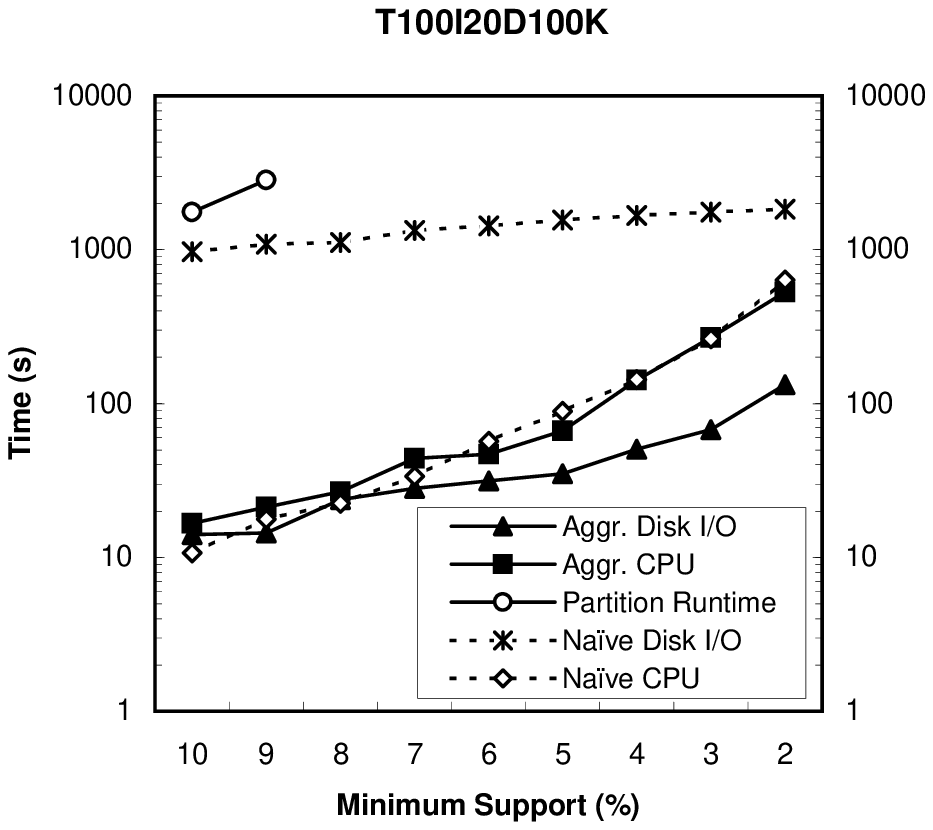,height=1.7in}}
       \center{\small (a)}
    \end{minipage}
    \hfill
    \begin{minipage}[t]{2in}
       \centerline{\psfig{figure=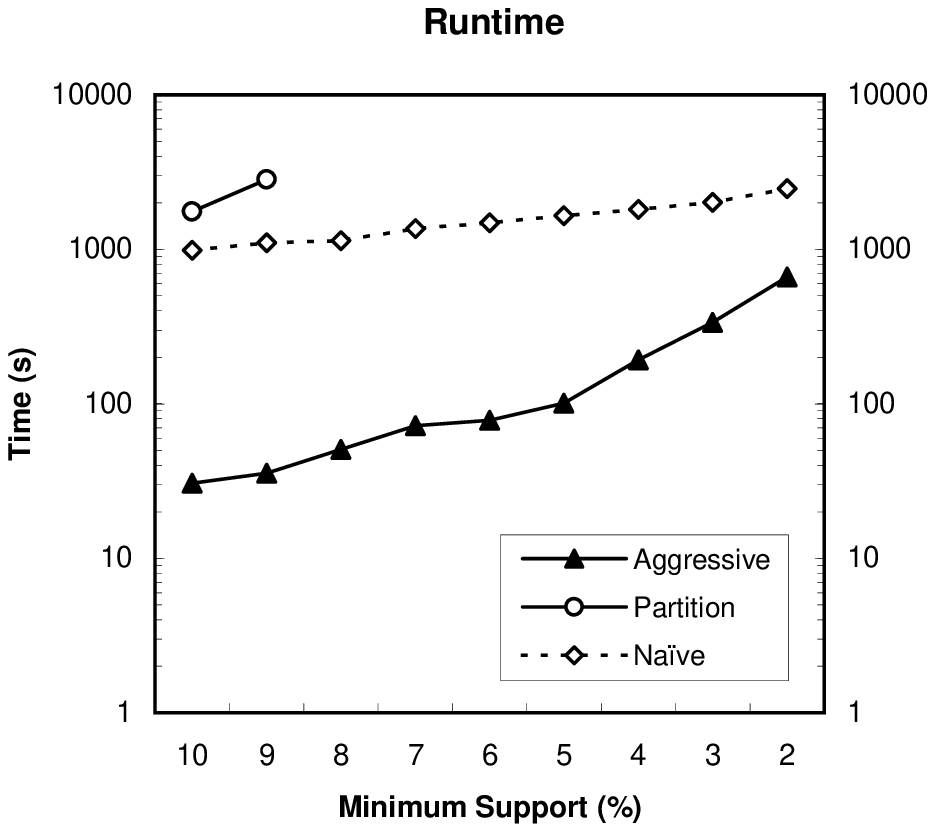,height=1.7in}}
       \center{\small (b)}
    \end{minipage}
    \hfill
    \begin{minipage}[t]{2in}
       \centerline{\psfig{figure=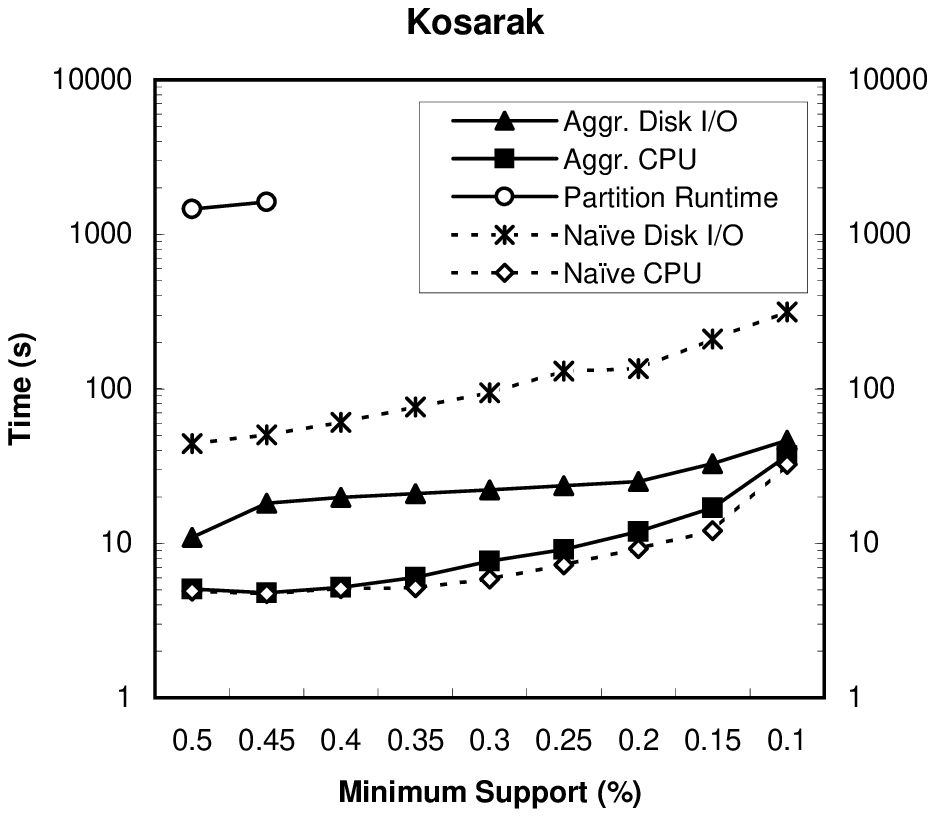,height=1.7in}}
       \center{\small (c)}
    \end{minipage}
  \caption{{\small Experiments on Synthetic Data and Real Data}}
  \label{SynReal}
\end{figure}

From \xf{SynReal} (a), 
we can see that the Partitioning Algorithm is the slowest is the group. 
The Naive Algorithm, however, is not slower than the Aggressive Algorithm 
if we only compare their CPU time. 
In \cite{fimi03}, 
where we concerned about main memory mining,
we found  that if a dataset is sparse the
boosted {\em FPgrowth*} method has a much better performance than
the original {\em FProwth}. 
The reason here the CPU time of the Aggressive Algorithm is not always
less than that of Naive Algorithm is
that the Aggressive Algorithm
has to spend CPU time on calculating statistics.
On the other hand, as expected,
we can see in the figure that
the disk I/O time of the Aggressive Algorithm is 
orders of magnitude smaller than that of the Naive Algorithm. 
In \xf{SynReal} (b) we compare the total runnng times.
We can see that the CPU overhead used by the Aggressive
Algorithm now become insignificant compared to
the savings in disk I/O.

We then ran the algorithms on a real dataset {\em Kosarak},
which is used as a test dataset in \cite{ZB03}.
The dataset is about 40 megabytes. 
Since it is a dense dataset and its FP-tree is pretty small,
we set the main memory size as 16 megabytes for the experiments.
Results are shown in \xf{SynReal} (c).

In \xf{SynReal} (b),
the Partitioning Algorithm is still the slowest.
This is  because it generates too many candidate frequent itemsets.
Together with the data structures, 
these candidate sets use up main memory and
virtual memory was used.
We can also again notice that the CPU time of the Naive Algorithm
is less than that of the Aggressive Algorithm.
This is because {\em Kosarak} is a dense dataset so
the array technique doesn't help a lot.
In addition, calculating the 
statistics takes much time.
The disk I/O's for the Aggressive Algorithm are still 
remarkably fewer than the disk I/O's for the Naive Algorithm.

To test the effectiveness of the techniques for grouping items,
we run {\em Diskmine} on 
{\em T100I20D100K} and see how 
close
the estimation of the FP-tree size for each group is to its real size.
We still set the main memory size as 128 megabytes, 
the minimum support is 2\%. 
When generating the projected databases, 
items were grouped into 7 groups 
(the total number of frequent items
is 826).
As we can see from \xf{Effect} (a),
in all groups, 
the estimated size is always slightly
than the real size. 
Compared with the Naive Algorithm, 
which constructs an FP-tree for each item from its projected database,
the Aggressive Algorithm almost fully 
uses the main memory for each group to
construct an FP-tree.   

\begin{figure}[ht!]
    \begin{minipage}[t]{1.5in}
       \centerline{\psfig{figure=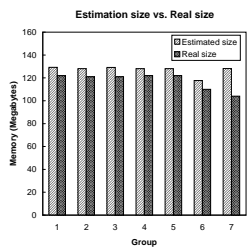,height=1.25in}}
       \center{\small (a)}
    \end{minipage}
    \hfill
    \begin{minipage}[t]{1.5in}
       \centerline{\psfig{figure=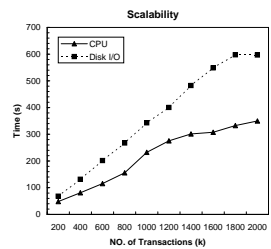,height=1.25in}}
       \center{\small (b)}
    \end{minipage}
  \caption{{\small Estimation Effect and Scalability of {\em Diskmine}}}
  \label{Effect}
\end{figure}
As a divide-and-conquer algorithm, 
one of the most important 
properties of {\em Diskmine} is its good scalability.
We ran {\em Diskmine} on a set of synthetic datasets.
In all datasets, 
the item number was set as 10000 items,
the average transaction length as 100,
and the average pattern length as 20.
The number of the transactions in the datasets 
varied from 200,000 to 2,000,000.
Datasets size ranges from 100 megabytes to 1 gigabyte.
Minimum support was set as 1.5\%, 
and the available main memory was 128 megabytes.
\xf{Effect} (b) shows the results.
In the figure, the CPU and the disk I/O time is 
always kept in a small range of acceptable values.
Even for the datasets with 2 million transactions,
the total running time is less than 1000 seconds.
Extrapolating from these figures using formula (4),
we can conclude that a dataset the size of the 
Library of Congress collection (25 Terabytes)
could be mined in around 18 hours with current technology.

\section{Conclusions}

We have introduced several divide-and-conquer algorithms 
for mining frequent itemset from secondary memory.
We have analyzed the 
recurrences and disk I/O's of all algorithms.

We then gave a detailed divide-and-conquer
algorithm 
which almost fully uses the limited main memory
and saves an numerous number of disk I/O's.
We introduced many novel techniques 
used in our algorithm.

Our
experimental results show
that our algorithm
successfully reduces the number of disk access,
sometimes by orders of magnitude,
and that our algorithm scales up to
terabytes of data.
The experiments also validates that
the estimation techniques used in 
our algorithm are accurate.

For future work,
we notice that 
there are very few efficient algorithm
for mining 
{\em maximal} frequent itemsets and {\em closed} 
frequent itemsets \cite{PBT99, PHM00,WHP03,Zaki02}
from very large databases.
Unlike in {\em Diskmine},
where the frequent itemsets mined from all projected databases
are globally frequent,
a maximal frequent itemset or a
closed frequent itemset mined from a projected database
is only locally maximal or closed.
As a challenge,
a data structure, whose size may be also very big,
must be set for recording all already discovered
maximal or closed frequent itemsets.
We also notice that 
our implementation of the partitioning algorithm is
based on an existing Apriori implementation,
which is not necessary highly optimized.
As we know,
there are situations
when there are not
too many candidate itemsets in a database,
but the FP-tree constructed from the database is pretty big.
In this situation the
Partitioning Algorithm only needs two database scans 
and all frequent items can be nicely mined in main memory,
or with very little I/O for keeping the
candidate sets in virtual memory.
In this situation
{\em Diskmine} also needs two database scans,
and it additionally
needs to 
decompose the database. 
Therefore, exploring whether some clever disk-based datastructure
would make the partition approach scale,
is another interesting direction for further research.

\end{document}